\documentclass[aps, preprint, superscriptaddress, nofootinbib]{revtex4}

\usepackage[usenames]{color}
\usepackage{graphicx}
\usepackage{amsmath}
\usepackage{amsfonts}
\usepackage{amssymb}
\usepackage{mathrsfs}
\usepackage{bm}
\usepackage{verbatim}
\usepackage{cancel}
\usepackage{enumitem}

\newcommand{\exLf}{\Psi}
\newcommand{\Lmbdstar}{\Lambda_{c}^\star}
\newcommand{\newSigma}{\Sigma(\pi \pi \Sigma_c, \frac{1}{2})}

\preprint{CTP-SCU/2016010}

\graphicspath{../}

\begin{document}

\title{Three-body system of $\pi \pi \Sigma_c$}

\author{Bingwei Long}
\email{bingwei@scu.edu.cn}
\affiliation{Center for Theoretical Physics, Department of Physics, Sichuan University, 29 Wang-Jiang Road, Chengdu, Sichuan 610064, China}

\date{October 17, 2016}

\begin{abstract}
The existence of near-threshold charmed baryon $\Lambda_c(2595)^+$ implies that the pion and the lightest, isospin-$1$ charmed baryon $\Sigma_c$ interact very strongly at extremely low energies. Using the two-flavor version of heavy hadron chiral perturbation theory, I explore the direct consequences of this strong force by investigating whether the $\Sigma_c$ can trap two very soft pions to form any visible hadronic states. The answer is positive. It is found without tuning any free parameters or ultraviolet cutoff that the state in question, with quantum numbers $I(J^P) = 1({\frac{1}{2}}^+)$, presents itself as a resonance pole only a few MeVs away from the $\pi \pi \Sigma_c$ threshold. Subleading corrections are estimated with power-counting arguments, and the smallness of pion momenta is found to facilitate the reliability of the analysis. Because of its proximity in mass, this excited $\Sigma_c$ resonance is speculated to be related to the broad resonance labeled as $\Lambda_c^+(2765)$.
\end{abstract}

\maketitle

Negative-parity, isoscalar, and spin-$1/2$ charmed baryon $\Lambda_c(2595)^+$ is situated $\delta \equiv \Delta - m_\pi \simeq 1$ MeV above the $\pi \Sigma_c$ threshold, where $\Sigma_c$ is the lightest isospin-$1$ charmed baryon, with a mass smaller than $\Lambda_c(2595)^+$ by $\Delta \simeq 139$ MeV~\cite{Agashe-2014kda}. Due to its quantum numbers, $\Lambda_c(2595)^+$, denoted by $\Lmbdstar$, couples in the $S$ wave to the isoscalar channel of $\pi \Sigma_c$. The closeness of $\Lmbdstar$ to the $\pi \Sigma_c$ threshold indicates that the resonant, $S$-wave interaction of $\pi \Sigma_c$ is incredibly strong at very low energies that are characterized by the size of the pion three-momentum around the $\Lmbdstar$ resonance, $Q \sim \sqrt{2 \delta m_\pi} \simeq 20$ MeV. This is an extreme case where the low-lying resonance upsets the naive expectation based on spontaneous breaking of the approximate chiral symmetry of QCD, that soft pions couple weakly to other hadrons.

While one can study the origin of this low-energy attraction of the $\pi \Sigma_c$ system per se, I explore here the direct consequences of this strong force, by investigating whether the $\Sigma_c$ can trap two soft pions to form heavier hadronic molecules. The $\pi \pi \Sigma_c$ configurations compatible with the isoscalar, $S$-wave $\pi \Sigma_c$ interaction have quantum numbers $I(J^P) = 1(\frac{1}{2}^+)$. Without tuning any free parameters at leading order (LO), a $\Sigma_c$ resonance is found to be near the $\pi \pi \Sigma_c$ threshold, with a pole at most a few MeVs away from the threshold.

The low-energy character of the $\pi \Sigma_c$ interaction makes it possible to focus on the small phase space around the $\pi \pi \Sigma_c$ threshold in which all three particles have momenta comparable to $Q \sim 20$ MeV. The $\Sigma_c$ can decay into the ground-state charmed baryon $\Lambda^+_c$ but the width $\simeq 2$ MeV is small. So the $\Sigma_c$ is approximated here as a stable state. Its transition to the ground-state $\Lambda_c$ and the pion can be incorporated in subleading orders, as will be briefly discussed. Since the constituent particles are almost on-shell and quite stable, nonrelativistic few-body dynamics is adequate and the states far away from the $\pi \pi \Sigma_c$ threshold can be ``integrated out''. In order to exploit systematically the smallness of $Q$, I use a specialized version of heavy hadron chiral perturbation theory (HHChPT)~\cite{Wise-1992hn, Burdman-1992gh, Yan-1992gz, Cho-1994vg} that includes only light flavors of $u$ and $d$.

The study echoes the efforts to investigate trapping of pions or kaons in finite~\cite{Dyson-1964xwa, Gal-2014zia, Akaishi-2002bg, KanadaEn'yo-2008wm} or infinite nuclear matter~\cite{Migdal-1990vm, Kaplan-1986yq}. However, the $\pi \Sigma_c$ interaction presents a more realistic scenario for pions to be trapped in baryonic matter, because the small pion momentum suggests that the effective field theory depends very little on the short-range detail of QCD physics.

Another interesting aspect of the $\Lmbdstar - \pi \Sigma_c$ system is the fine-tuning manifested by the tiny value of $\delta$. A near-threshold $S$-wave resonance usually implies both the scattering length and effective range be fine-tuned, but Ref.~\cite{Long-2015pua} was able to show that thanks to chiral symmetry, only one fine-tuning, the pion mass, is needed for the underlying theory to situate the $\Lmbdstar$ so close to the $\pi \Sigma_c$ threshold. Looking into how this fine-tuning propagates through the charmed-baryon sector provides a perspective that could offer more insights into hadronic interactions. For instance, the decay phenomenology of $\Lmbdstar$ becomes very sensitive to isospin violations~\cite{Blechman-2003mq, Guo-2016wpy} due to the smallness of $\delta$. The three-body system of $\pi \pi \Sigma_c$ is another natural stage to look for the implication of the said fine-tuning, in a spirit similar to studying universality in few-body systems with large scattering length~\cite{Braaten-2004rn}.

For the time being, only $\Sigma_c$, $\Lmbdstar$ and pions are relevant degrees of freedom, so we consider the usual heavy-baryon chiral Lagrangian without heavy quark symmetry manifestly incorporated. The relevant leading terms~\cite{Long-2015pua} are
\begin{equation}
    \begin{split}
    \mathcal{L} &= i\Sigma_a^\dagger \dot{\Sigma}_a + \frac{i}{f_\pi^2} \Sigma_a^\dagger \left( \pi_a \dot{\pi}_b - \pi_b \dot{\pi}_a \right) \Sigma_b \\
    &\;\; + \exLf^\dagger \left( i\partial_0 - \Delta \right) \exLf + \frac{h}{\sqrt{3} f_\pi} \left( \Sigma_a^\dagger \dot{\pi}_a \exLf + \text{H.c.} \right) \\
    &\;\; + \left(\frac{m_\pi^2}{2}  \bm{\pi}^2 - \dot{\bm{\pi}}^2 \right)\frac{\bm{\pi}^2}{4f_\pi^2}
    + \cdots
    \label{eqn_nu0}
    \end{split}
\end{equation}
Here $\exLf$ ($\Sigma$) is the field that annihilates $\Lmbdstar$ ($\Sigma_c$).  The pion decay constant $f_\pi = 92.4$ MeV, and the $\pi \Sigma_c \Lmbdstar$ transition coupling $h^2 = 3/2\, h_2^2$, where $h_2$ is the counterpart of $h$ in the HHChPT Lagrangian~\cite{Cho-1994vg}. At LO, the transition vertex is approximately proportional to $m_\pi$ because the pion momenta are very small. The second term of the first line is the Weinberg-Tomazawa term for the $\Sigma_c$, and the third line is the leading $S$-wave pion-pion interaction. I use throughout the paper the heavy-baryon notation for baryon energies, which have the mass of $\Sigma_c$ subtracted.

The two-body interaction of $\pi \Sigma_c$ is can be encapsulated in the dressed $\Lmbdstar$ propagator~\cite{Long-2015pua},
\begin{equation}
    i D(p) = \frac{i}{(p_0 - m_\pi) -\delta - \epsilon h^2 \sqrt{2(m_\pi - p_0) m_\pi}} \, ,
    \label{eqn_Lmbdpropagator}
\end{equation}
where, besides $\delta$, $\epsilon \equiv m_\pi^2/4\pi f_\pi^2 = 0.18$ is the other small parameter to be exploited here. Since we are only interested in very low energies at which $\pi \Sigma_c$ interaction is resonant, we can consider for power-counting purposes that $|p_0 - m_\pi| \sim \delta$. The dressing is necessary when each term of the denominator in Eq.~\eqref{eqn_Lmbdpropagator} is of the same size. It immediately follows that $\delta \sim Q^2/m_\pi \sim \epsilon^2 m_\pi$, where we have used $h = \mathcal{O}(1)$. It can be numerically verified that $\delta$ and $\epsilon^2 m_\pi$ are indeed of the same order of magnitude.

Attaching $\pi \Sigma_c \Lmbdstar$ vertexes to the dressed $\Lmbdstar$ propagator, we obtain the $\pi \Sigma_c$ elastic scattering amplitude and extract
the scattering length and effective range as follows~\cite{Long-2015pua},
\begin{equation}
    -1/a = \frac{\delta}{\epsilon h^2}, \quad r = - \left( \epsilon h^2 m_\pi \right)^{-1} \, .
    \label{eqn_ar}
\end{equation}

\begin{figure}
    \centering
    \includegraphics[scale=0.45]{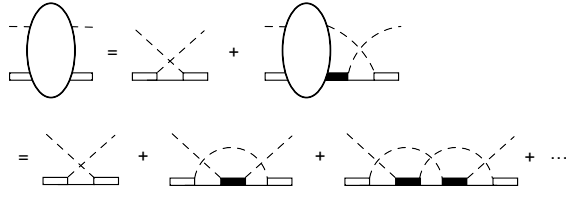}
    \caption{Resummation of $\Sigma_c$-exchanges in $\pi \Lmbdstar$ scattering. The double, solid, and dashed lines represent propagation of a $\Lmbdstar$, a $\Sigma_c$, and a pion, respectively. The thick lines are dressed $\Lmbdstar$ propagator.}
    \label{fig_blob}
\end{figure}

Since there is not any physical difference between the $\Psi$ field and composite operator $\pi_b \Sigma_b$, any correlation functions of the form $\langle 0| \pi_a \Psi \pi_a \Psi^\dagger |0 \rangle$ can be used to search for potential states associated with $\pi \pi \Sigma_c$. I choose to study the pole structure of the $\pi \Lmbdstar$ scattering amplitude, represented by the blob in Fig.~\ref{fig_blob}. In the center-of-mass (CM) frame, the pion has incoming (outgoing) four-momentum $(k_0 + m_\pi, \vec{k})$ [$(q_0 + m_\pi, \vec{q}\,)$] and the baryon has incoming (outgoing) four-momentum $(E_\Lambda + m_\pi, -\vec{k})$ [$(E - q_0 + m_\pi, -\vec{q}\,)$], where $E_\Lambda$ is the energy of $\Lmbdstar$. In my notation the CM energy $\sqrt{s} = E + 2m_\pi + M_{\Sigma_c}$, and $E = \vec{q}\,^2/2m_\pi + E_\Lambda$ when the external pions are on-shell, but the external $\Lmbdstar$ are not necessarily so.

We can break up any $\pi \Lmbdstar$ scattering diagrams into two parts: (1) $\pi \Lmbdstar$ potentials, diagrams that are still connected after a pion and a $\Lmbdstar$ internal lines are cut, and (2) propagation of $\pi \Lmbdstar$ with the dressed $\Lmbdstar$ propagator. The dominant $\pi \Lmbdstar$ potential is the $u$-channel $\Sigma_c$-exchange. Illustrated in the second line of Fig.~\ref{fig_blob} are $\Sigma_c$ exchanges connected by $\pi \Lmbdstar$ propagators. Using power-counting language, I argue as follow that these diagrams must be resummed.

The pion's kinetic energy is $\sim Q^2/m_\pi$, so is the energy following through baryon propagators. Therefore, the $\Sigma_c$ propagator contributes a factor of $(Q^2/m_\pi)^{-1}$. With the $\pi \Sigma_c \Lmbdstar$ vertex $\sim (m_\pi/f_\pi)$, the LO potential is then counted as
\begin{equation}
   \frac{m_\pi}{f_\pi} \frac{1}{Q^2/m_\pi} \frac{m_\pi}{f_\pi} \sim \frac{m_\pi^3}{f_\pi^2 Q^2} \, .
\end{equation}

The propagation of $\pi \Lmbdstar$ intermediate states consists of a pion propagator contributing a factor of $1/Q^2$, a dressed $\Lmbdstar$ propagator $\sim (Q^2/m_\pi)^{-1}$, and the loop integration $\int dl_0 d^3 l \sim (Q^2/m_\pi) Q^3$. The numerical factor associated with nonrelativistic pion loops is typically $1/(4\pi)$, rather than $1/(16\pi^2)$~\cite{Long-2015pua}. Therefore, the $\pi \Lmbdstar$ propagation generally contributes a factor of $Q/4\pi$.

The once-iterated potential, the second diagram in the second line of Fig.~\ref{fig_blob}, scales as
\begin{equation}
    \frac{m_\pi^3}{f_\pi^2 Q^2}\, \frac{Q}{4\pi}\, \frac{m_\pi^3}{f_\pi^2 Q^2} \sim \frac{m_\pi^3}{f_\pi^2 Q^2} \frac{Q}{\epsilon m_\pi}
\end{equation}
Because $Q/\epsilon m_\pi \sim 1$, the once-iterated potential contributes about the same as the Born term does. By induction, we conclude that it is necessary to resum all the diagrams in the second line of Fig.~\ref{fig_blob}.

\begin{figure}
    \centering
    \includegraphics[scale=0.5]{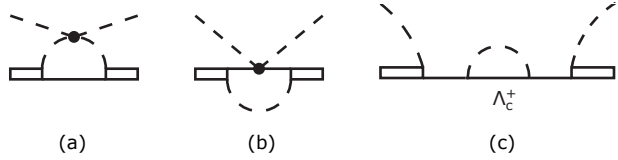}
    \caption{Subleading $\pi \Lmbdstar$ potentials. Except that the baryon propagator at the center of (c) represents the ground-state $\Lambda_c^+$, the symbols are the same as in Fig.~\ref{fig_blob}.}
    \label{fig_nlo}
\end{figure}

With the above argumentation, we are in a position to estimate theoretical uncertainties of the present analysis by counting subleading corrections that are not included at LO. In Fig.~\ref{fig_nlo} (a) $S$-wave pion-pion vertexes contribute a factor of $m_\pi^2/f_\pi^2$. With the aforementioned counting rule applied to other elements of the diagram, pion-pion interactions are found to correct the LO $\pi \Lmbdstar$ potential by $\mathcal{O}(\epsilon^2)$. However, this does not mean that the problem reduces to a simple system of independent bosons, because the energy-dependent $\pi \Sigma_c$ invalidates the separation of the pions' coordinates in the Schr\"odinger equation.

The Weinberg-Tomozawa term for the $\Sigma_c$ provides corrections to both the $\Lmbdstar$ self energy and the $\pi \Lmbdstar$ potential. The diagram of Fig.~\ref{fig_nlo} (b) shows its correction to the potential, but it turns out to vanish after the isospin indexes are contracted. Even if it did not, its contribution would be $\mathcal{O}(\epsilon^2)$, based on power counting. The correction to the $\Lmbdstar$ self energy was found in Ref.~\cite{Long-2015pua} to be $\mathcal{O}(\epsilon^2)$.

Since the $\Sigma_c$ couples to $\pi \Lambda_c^+$, where $\Lambda_c^+$ is the ground-state charmed baryon, it is necessary to analyze the contribution of the $\pi \Lambda_c^+$ intermediate states. Figure~\ref{fig_nlo} (c) shows how the $\pi \Lambda_c^+$ intermediate states contribute. The four-momentum flowing through the loop is of the size $Q' \sim (M_{\Sigma_c} - M_{\Lambda_c^+} + 2m_\pi)$, which can be numerically approximated by $\sim 3 m_\pi$; therefore, we can apply the standard ChPT counting to relativistic pions. The $\Sigma_c$ propagators are off-shell by an amount of $2m_\pi$, and the $\pi \Lambda_c^+ \Sigma_c$ transition vertex contributes a factor of $Q'$. Putting these elements together, we find the diagram scales
\begin{equation}
    \left(\frac{m_\pi}{f_\pi}\right)^2 \left(\frac{1}{2m_\pi}\right)^2 \frac{{Q'}^3}{(4\pi f_\pi)^2} \, ,
\end{equation}
which is suppressed by a factor of $\frac{3}{2} \epsilon^2 (\frac{Q'}{4\pi f_\pi})^2$, compared with the LO potential. Intermediates states involving more pions, like $\pi \pi \Lambda_c^+$, give rise to even more loops, and hence are more suppressed.

$\Sigma_c$ and $\Lmbdstar$ have respectively a spin-$3/2$ neighboring state, $\Sigma_c(2520)$ and $\Lambda_c^+(2625)$. These neighboring states are degenerate in the heavy quark limit. If we only search for possible resonances with $I(J^P) = 1 ({\frac{1}{2}}^+)$, these spin-$3/2$ partners will not interfere very much despite the relatively small mass difference. For example, consider the $u$-channel exchange between $\Lmbdstar$ and $\pi$ by a $\Sigma_c(2520)$, denoted by $\Sigma_c^\star$. The $\pi \Sigma_c^\star \Lambda_c^\star$ transition vertex is proportional to the pion momentum square~\cite{Cho-1994vg}. After being projected onto the $S$-wave, the $\Sigma_c^\star$ exchange is suppressed by $\mathcal{O}(\epsilon^2 Q^2/M_{hi}^2)$, where $M_{hi}$ is the break-down scale of this specialized, two-flavor ChPT.

The above power counting applies to expansion of the amplitude, not that of properties like the pole position that are extracted from the amplitude. While the LO amplitude will establish qualitatively the existence of the advocated resonance, the LO value of the imaginary part of its pole position does not necessarily reflect the total decay width, at least not to the accuracy level discussed above.

Let us proceed to more quantitative analyses. With $T(\vec{q}; E, E_\Lambda, q_0)$ representing the $\pi \Lmbdstar$ amplitude, the first line of Fig.~\ref{fig_blob} translates into the following integral equation,
\begin{equation}
    \begin{split}
    & T(\vec{q}; E, E_\Lambda, q_0) = - \frac{h^2 m_\pi^2}{3 f_\pi^2 (E_\Lambda - q_0 + i0)}\\
    & + i h^2 \frac{ m_\pi^2}{3 f_\pi^2} \int \frac{d^4 l}{(2\pi)^4} \frac{1}{E - q_0 - l_0 + i0} \frac{1}{2 m_\pi l_0 - \vec{l}\,^2 + i0} \\
    & \times \frac{T(\vec{l}; E, E_\Lambda, l_0)}{E - l_0 - \delta - \epsilon h^2 \sqrt{-2(E - l_0) m_\pi - i0} + i0} \, ,
    \end{split}
\end{equation}
where $l_0^2$ has been dropped off in the pion propagator, since $l_0 \sim {\vec{l}}^2/2m_\pi$. Integrating over $l_0$ and the angular part of $\vec{l}$, setting $q_0 = \vec{q}^2/2m_\pi$ to define $t(q; \mathcal{E}, \mathcal{B}) \equiv T(\vec{q}; E, E_\Lambda, q^2/2m_\pi)$, we arrive at
\begin{equation}
\begin{split}
    & t(q; \mathcal{E}, \mathcal{B}) = \frac{8\pi/|r|}{3(q^2 + \mathcal{B})} + \frac{2}{3 \pi} \int_{\Sigma_l} dl \frac{l^2}{q^2 - \mathcal{E} + l^2 + i0} \\
    & \quad \times \frac{t(l; \mathcal{E}, \mathcal{B})}{-\frac{1}{a} - \frac{|r|}{2}(\mathcal{E} - l^2) + \sqrt{l^2 - \mathcal{E} - i0} -i0} \, , \label{eqn_1dinteqn}
\end{split}
\end{equation}
where $\mathcal{E} \equiv 2m_\pi E$, $\mathcal{B} \equiv -2m_\pi E_\Lambda$, and the $\pi \Sigma_c$ scattering length and effective range have been used to make the notation more compact. The subscript $\Sigma_l$ serves to remind that in order to continue $t(q; \mathcal{E}, \mathcal{B})$ to the complex $\mathcal{E}$ plane, we need to deform the integration contour away from the positive real axis. Since we are only interested in extracting the pole position, the field renormalization constants of $\pi$ and $\exLf$ are not accounted for.

The integral in Eq.~\eqref{eqn_1dinteqn} actually converges. To see this, note that the $q$ dependence on the right hand side suggests that when $q \to \infty$, $t(q; \mathcal{E}, \mathcal{B})$ vanishes as fast as $q^{-2}$. The convergence is also confirmed numerically. It is rather important that the pole position extracted is independent of the way the integral is regularized, for we can then state with confidence that the sought-after hadronic structure does not come out of modeling short-range QCD physics.

In order to continue analytically the above integral equation into the complex $\mathcal{E}$ plane, one must deform tactfully the integration contour so that, as $\mathcal{E}$ moves into its second sheet, it does not interfere any singularities of the integrand. The technique used here is similar to that of Ref.~\cite{Pearce-1984ca}, i.e., rotating the $l$ contour $l \to l e^{-i\phi}$. Reference.~\cite{Pearce-1984ca} accounted for the singularities of two propagators as functions of $l$, but did not discuss the possible $l$-singularities of $t(l; \mathcal{E}, \mathcal{B})$ as a function of $l$. Remarkably, even after taking into consideration the singularities of $t(l; \mathcal{E}, \mathcal{B})$ as a function of $l$, one can show that the prescription of Ref.~\cite{Pearce-1984ca} does not need to change. The technical details of the calculation is carried out will be shown in a later publication~\cite{LvLongOnGoing}.

\begin{figure}
    \centering
    \includegraphics[scale=0.6]{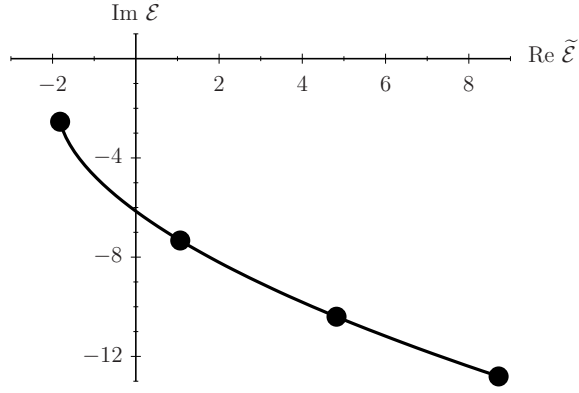}
    \caption{As $\tilde{\delta}$ changes, the pole trajectory of the $\newSigma$. From left to right, the filled circles correspond to $\tilde{\delta} = -1, -2, -3, -4$, respectively. (See the text for more detailed definitions of $\widetilde{\mathcal{E}}$ and $\tilde{\delta}$.)}
    \label{fig_polepos}
\end{figure}

Numerical calculations indeed indicate that there exists a resonance state with $I(J^P) = 1 ({\frac{1}{2}}^+)$, situated near the $\pi \pi \Sigma_c$ threshold. I denote this state by $\newSigma$ in the present paper. Its existence is manifested by the resonance pole of $\pi \Lmbdstar \to \pi \Lmbdstar$ amplitude. A mathematically compact way to present the pole position is to define dimensionless quantities
\begin{equation}
    \widetilde{\mathcal{E}} \equiv \mathcal{E}/(\epsilon h^2 m_\pi)^2, \quad \tilde{\delta} \equiv \delta/(\epsilon^2 h^4 m_\pi)\, ,
\end{equation}
and to show how the pole position in the $\widetilde{\mathcal{E}}$ plane varies with $\tilde{\delta}$, $\widetilde{\mathcal{E}}_\text{pole} = \widetilde{\mathcal{E}}_\text{pole}(\tilde{\delta})$. Figure~\ref{fig_polepos} shows the pole trajectory as $\tilde{\delta}$ varies from $-4$ to $-1$.

Because $\Lmbdstar$ is only a couple of MeVs away from the $\pi \Sigma_c$ threshold, its properties $h^2$ and $\delta$, determined from the pionic decay data, are sensitive to the mass splitting between $\pi^0$ and $\pi^\pm$. Before a more accurate calculation is carried out~\cite{LvLongOnGoing}, we can have a flavor of the pole position of $\newSigma$ by applying two sets of parameters to the present isospin-invariant calculation, with the isospin-averaged pion and $\Sigma_c$ masses adopted, $m_\pi = 138.0$ MeV and $M_{\Sigma_c} - M_{\Lambda^+_c}= 167.1$ MeV. One has a higher $\Lmbdstar$ mass~\cite{Chiladze-1997ev}:
\begin{equation}
    M_{\Lmbdstar} - M_{\Lambda^+_c} = 308.7\, \text{MeV}\, , \quad h^2 = \frac{3}{2} \times 0.30 \, ,
\end{equation}
which gives $\newSigma$ the following pole position,
\begin{equation}
M_{\newSigma} - (M_{\Sigma_c} + 2m_\pi) = (4.00 - 5.72i) \text{MeV}\, .
\end{equation}
The other is from Ref.~\cite{Aaltonen-2011sf}:
\begin{equation}
    M_{\Lmbdstar} - M_{\Lambda^+_c} = 305.8\, \text{MeV}\, , \quad h^2 = \frac{3}{2} \times 0.36 \, ,
\end{equation}
resulting in the pole being situated slightly below the $\pi \pi \Sigma_c$ threshold,
\begin{equation}
M_{\newSigma} - (M_{\Sigma_c} + 2m_\pi) = (-0.45 - 0.02i) \text{MeV}\, .
\end{equation}

If we replace $\Sigma_c$ and $\Lmbdstar$ with their spin-$3/2$ partners, $\Sigma_c(2520)$ and $\Lambda^+_c(2625)$, and repeat the above analysis, it is likely to find the spin-$3/2$ partner of the $\newSigma$, with a mass a few tens of MeVs heavier. If this turns to be the case, it is conceivable to identify the pair with the lower broad peak, labeled by $\Lambda^+_c(2765)$ in Ref.~\cite{Agashe-2014kda}, and observed by CLEO in decays into $\Lambda^+_c \pi^- \pi^+$~\cite{Artuso-2000xy} where it was not ruled out that the peak could be two overlapping states. Studies based on quark models related to $\Lambda^+_c(2765)$ can be found in, for examples, Refs.~\cite{Capstick-1986bm, Copley-1979wj}.

\begin{figure}
    \centering
    \includegraphics[scale=0.6]{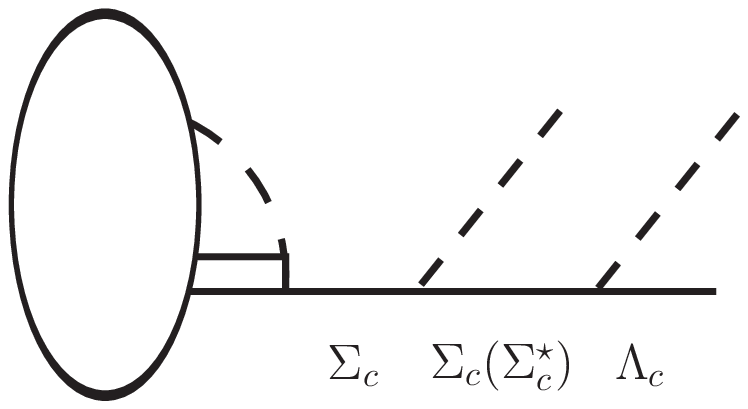}
    \caption{The decay of $\newSigma$ into $\Lambda^+_c \pi^- \pi^+$. The blob represents the composite structure of $\newSigma$.}
    \label{fig_decay}
\end{figure}

While a more careful confrontation with the invariant mass spectrum data is underway~\cite{LvLongOnGoing} to determine whether $\Lambda_c^+(2765)$ is indeed $\newSigma$ or the overlapping of $\newSigma$ and its spin-$3/2$ partner, I point out here that the decay of $\newSigma$ into $\Lambda^+_c \pi^- \pi^+$ is possible, with the dominant contribution illustrated in Fig.~\ref{fig_decay}. From left to right, the first solid line represents a $\Sigma_c$ intermediate state that is above its energy shell by about $2m_\pi$. After emitting a pion, the $\Sigma_c$ could become on-shell, either remaining to be itself or turning into $\Sigma_c(2520)$. Then the $\Sigma_c$ or $\Sigma_c(2520)$ decays into $\Lambda^+_c \pi$. This decay mechanism is consistent with the finding of Ref.~\cite{Artuso-2000xy} that $\Lambda^+_c(2765)$ appears to resonate through $\Sigma_c$ and probably also $\Sigma_c^\star$.

\acknowledgments

I thank Fei Huang for useful discussions about the analytic continuation technique used in Ref.~\cite{Doring-2009yv} and Songlin Lv for finding the relation between $h$ and $h_2$ in the Lagrangian~\eqref{eqn_nu0}. Also acknowledged is the hospitality of the nuclear theory group at BeiHang University where part of the work was done. This work was supported in part by the National Natural Science Foundation of China (NSFC) under Grant No. 11375120.

\end{document}